\documentclass[]{spie}  

\usepackage{amsmath,amsfonts,amssymb}
\usepackage{graphicx}
\usepackage[colorlinks=true, allcolors=blue]{hyperref}
\usepackage{comment}
\usepackage{subcaption}

\title{Motorized Mount for the 3-DoF Folding Mirror 2 of the VLT's BlueMUSE Instrument}

\author[a]{Gloria Mellinand}
\author[a]{Diane Chapuis}
\author[a]{Malak Galal}
\author[a]{Eirini Tagkoudi}
\author[a]{Evan Touraine}
\author[a]{Sébastion Pernecker}
\author[b]{Rémi Giroud}
\author[b]{Alexandre Jeanneau}
\author[b]{Florence Laurent}
\author[b]{Johan Richard}
\author[c]{Chris Coote}
\author[c]{Jon Moller}
\author[a]{Jean-Paul Kneib}
\affil[a]{Institute of Physics, Laboratory of Astrophysics, Ecole Polytechnique Fédérale de Lausanne (EPFL), Observatoire de Sauvergny, CH-1290 Versoix, Switzerland}
\affil[b]{Univ Lyon, Univ Lyon1, Ens de Lyon, CNRS, Centre de Recherche Astrophysique de Lyon UMR5574, F-69230,
Saint-Genis-Laval, France}
\affil[c]{Australian Astronomical Optics, Macquarie University, NSW, Australia}

\authorinfo{Further author information: (Send correspondence to G.M., D.C. and M.G.)\\G.M.: E-mail: gloria.mellinand@epfl.ch, Telephone: +41 77 218 78 09\\ D.C.: E-mail: diane.chapuis@epfl.ch, Telephone: +41 21 693 54 37\\ M.G.: E-mail: malak.galal@epfl.ch, Telephone: +41 21 693 92 91}

\begin{document} 
\maketitle

\begin{abstract}
BlueMUSE is a blue-optimized, medium spectral resolution, panoramic integral field spectrograph under development for the Very Large Telescope (VLT). The project is now fully entering design phase. With an optimized transmission down to 350 nm, spectral resolution of R~3500 on average across the wavelength range, and a large FoV (1 arcmin²), BlueMUSE will open up a new range of galactic and extragalactic science cases facilitated by its specific capabilities.

To meet the stability demands required for BlueMUSE, motorized mounts for precise and repeatable positioning of key optics are developed. This paper explores candidate mechanical designs for BlueMUSE's Folding Mirror 2 (FM2), a high-precision mirror mount with three degrees-of-freedom: tip, tilt rotations, and vertical translation. The study includes a comparative performance analysis through theoretical simulations, details the mechanical, software and electronics architecture of the chosen design, as well as the dedicated optical setup design to characterize repeatability, precision, and thermal stability. The resulting performance of the chosen FM2 mount is then evaluated against the specified requirements for BlueMUSE.
\end{abstract}

\keywords{Motorization, Folding Mirror, BlueMUSE, VLT}

\section{INTRODUCTION}
\label{sec:intro}  

\begin{figure}[!h]
    \centering
    \includegraphics[scale=0.4]{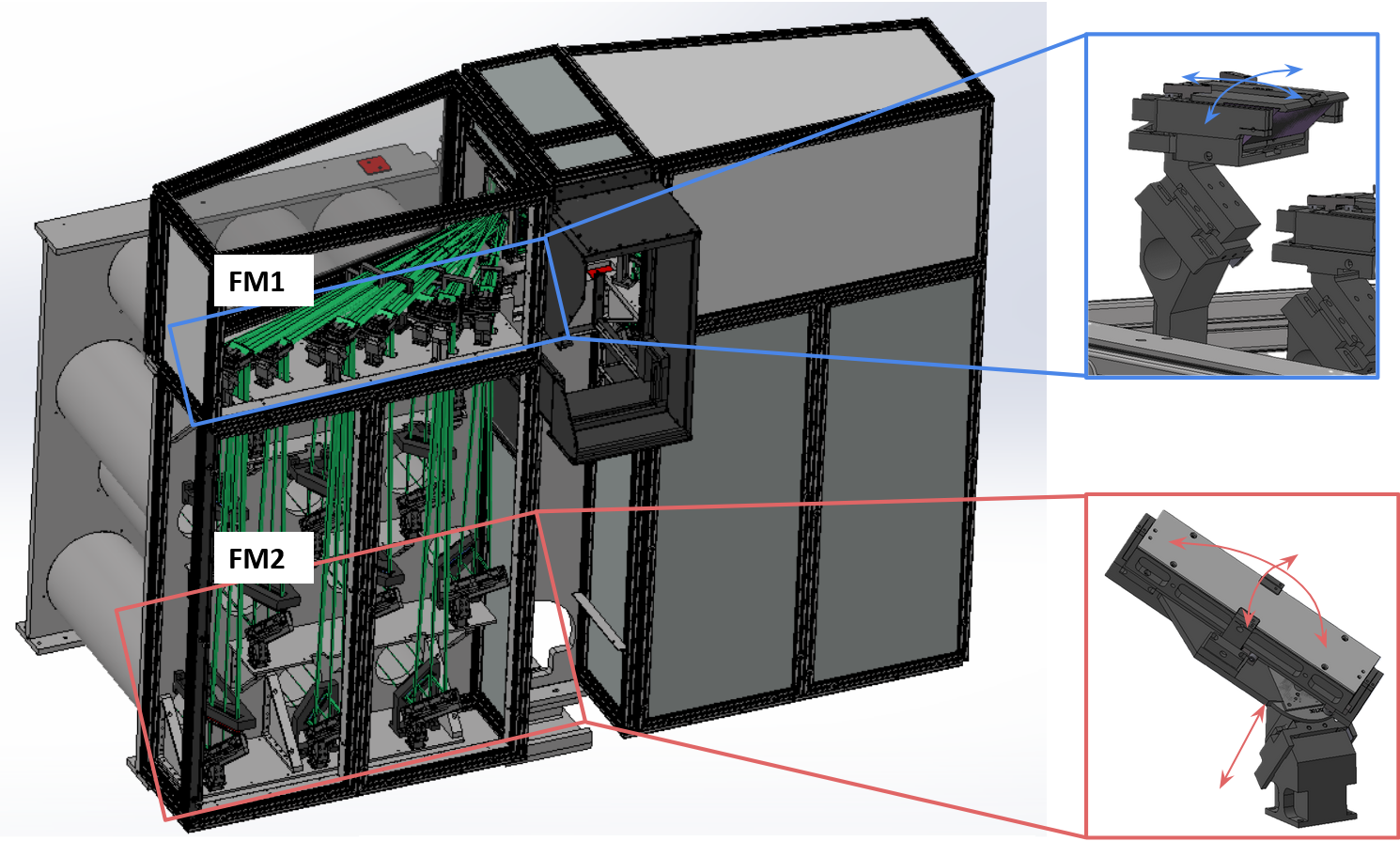}
    \caption{CAD view of the MUSE instrument and the Folding Mirrors (FM) within. The optical paths are shown in green, and the left hand side of the instrument is mirrored on the right. Blue arrows show the FM1's rotational degrees of freedom, while red arrows show the FM2's addition of a translational degree of freedom}  
    \label{fig:MUSE_FM}
\end{figure}

BlueMUSE is a next-generation integral field spectrograph currently under development for installation on the Very Large Telescope (VLT) \cite{Richard2021BlueMUSE,Richard2024BlueMUSE}. Designed to operate in the blue region of the visible spectrum, the instrument will provide high-resolution spectroscopic imaging capabilities for a broad range of scientific applications, including the study of young stellar populations, galaxy evolution, and faint diffuse structures in the universe. To achieve these objectives, BlueMUSE makes use of a complex optical architecture composed of custom blue-optimized spectrographs and an extensive relay optics system\cite{jeanneau2020curved}. The instrument builds directly upon the successful MUSE concept \cite{Bacon2010MUSE,Bacon2015MUSE}, while introducing more demanding optical requirements driven by its shorter operating wavelengths.

Meeting these requirements relies on maintaining the alignment of numerous optical components distributed throughout the instrument. Among these, the Folding Mirrors (FMs) play a critical role in directing the optical beam through the relay optics chain. A total of 32 folding mirrors are implemented within BlueMUSE, divided into two distinct configurations. Folding Mirror 1 (FM1) provides two rotational Degrees of Freedom (DoF) corresponding to the tip ($\Theta Y_{FM}$) and tilt ($\Theta X_{FM}$) adjustments of the mirror surface \cite{Mellinand2026FM1}. Folding Mirror 2 (FM2) incorporates the same rotational motions together with an additional translational degree of freedom normal to the mirror surface. Sixteen units of each mirror type are distributed throughout the instrument.

The alignment stability of these mirrors is strongly influenced by environmental conditions. Although the VLT site at Cerro Paranal experiences relatively small annual temperature variations due to its stable arid climate \cite{sarazin1999climate}, temperature differences between daytime and nighttime operation are sufficient to induce thermo-mechanical deformation in precision opto-mechanical assemblies. Previous studies conducted on the MUSE structure have demonstrated the sensitivity of optical performance to such thermal effects \cite{Cai2024MUSEFEA}. Even small structural deformations can generate critical beam deviations, resulting in alignment drift and degradation of instrument performance \cite{NOETHE20021}\cite{fischer2008optical}.

The current MUSE folding mirror mounts rely on manual adjustment mechanisms. As a consequence, thermal-induced alignment variations occurring throughout the day and between exposures at night can require frequent recalibration by trained personnel\cite{Weilbacher2020data}. These procedures are both time-consuming and delicate, to ensure arcsecond-level beam stability is maintained. Reducing the need for manual intervention therefore represents an important advancement for the next generation of folding mirror systems.

These limitations motivate the development of compact motorized folding mirror platforms capable of active repositioning while preserving long-term optical stability. Achieving this objective presents several engineering challenges. Positioning accuracy can be affected by multiple sources of error, such as thermal expansion, structural compliance, assembly tolerances, and mechanical backlash. Ensuring arcsecond-level repeatability requires optimization of the mechanical architecture, material selection, and actuation strategy\cite{giesen2003design}. To evaluate the mirror mounts, dedicated metrology systems must be developed to quantify both repeatability and thermal stability at the required level of precision.

This paper presents the system architecture and two potential mechanical designs of the FM2 motorized mount, with a description of its two rotational axes and one translation axis. We then describe the dedicated experimental setup in development to evaluate the system according to repeatability and short and long-term thermal stability requirements. Finally, we discuss the theoretical results obtained with thermal simulations and outline future developments for the next iteration of the Folding Mirror 2 design.

\section{FM2 Mechanical Designs and Systems Architecture}
\begin{figure}[!h]
    \centering
    \begin{subfigure}[b]{0.40\textwidth}
        \includegraphics[width=\textwidth]{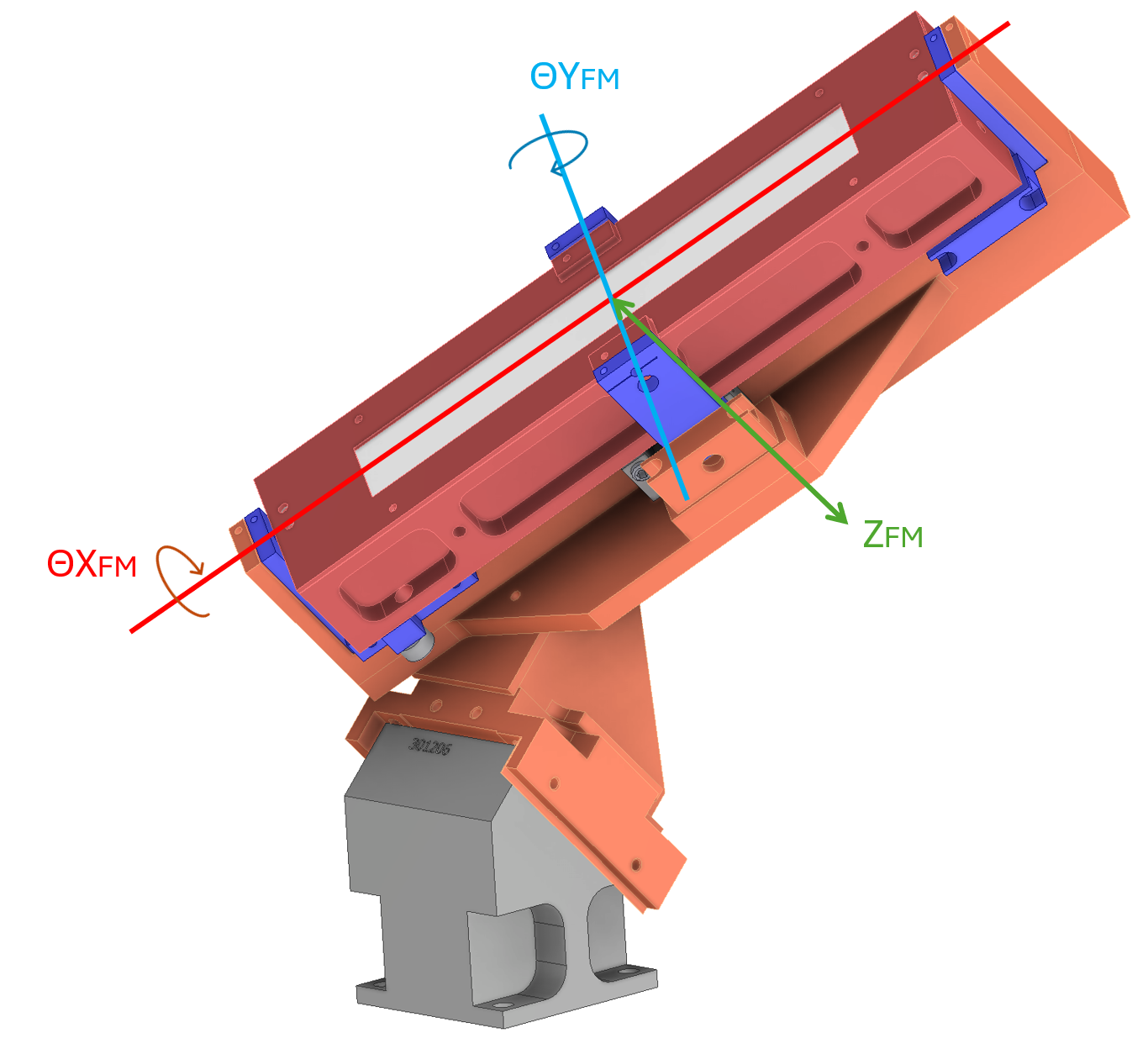}
        \caption{}  
        \label{fig:FM2_MUSE}
    \end{subfigure}
    \hfill
    \begin{subfigure}[b]{0.55\textwidth}
        \centering
        \includegraphics[width=\textwidth]{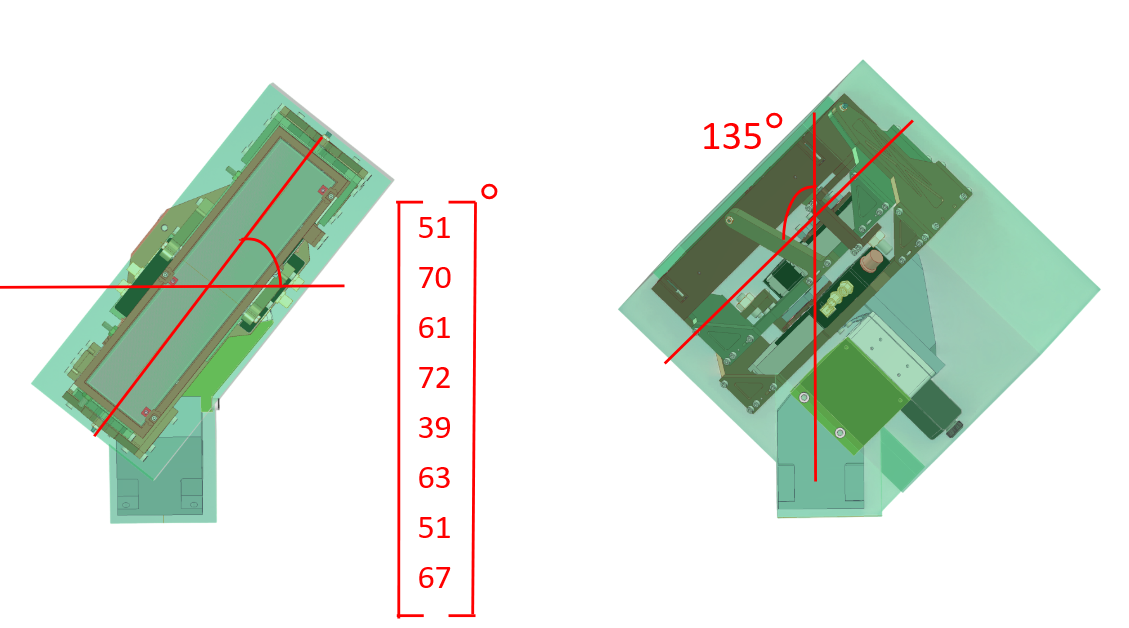}
        \caption{}  
        \label{fig:FM2_configs}
    \end{subfigure}
    \caption{Folding mirror 2 (FM2) overview: (a) The MUSE FM2 with 3 degrees of freedom; (b) FM2's configurations. Shown in green is the space enveloppe of the designed BlueMUSE FM2. In red are the required orientations of the mirror. 8 angles are defined for mirror orientation, due to the light path redirection inside of the instrument.} 
    \label{fig:FM1_new}
\end{figure}

The FM2 mechanism is actuable in rotation along the $\Theta X_{FM}$ and $\Theta Y_{FM}$ axis, and in translation along the $Z_{FM}$. The axes of rotation are shown in Figure \ref{fig:FM2_MUSE} and are aligned along the mirror surface, while the translation moves the mirror over a range of 2 mm perpandicular to the mirror surface.  

The desired performance defining the Folding Mirrors' design and performance describe repeatability, accuracy and range. The following table shows preliminary identified requirements for the Folding Mirror 2 (FM2) mount:

\begin{table} [!h]
\centering
\begin{tabular}{lccc}
\hline\hline
Requirement & FM2 $\Theta Y_{FM}$ & FM2 $\Theta X_{FM}$ & FM2 $Z_{FM}$ \\
\hline
Repeatability in one step & 7.5 arcsec & 7.5 arcsec & 35 $\mu m$\\
Mount stability during operation & 1.5 arcsec & 1.5 arcsec & 7.5 $\mu m$  \\
Long-term mount stability & 1.5 arcsec & 1.5 arcsec & 7.5 $\mu m$  \\
Range & 10 arcmin & 10 arcmin & 2 mm\\
\hline\end{tabular}
    \caption{FM2 motorized mount performance requirements updated in January 2026}
    \label{tab:FM2_req}
\end{table}

The mount stability requirement defines short term stability of the mirror mount during 12h, and is justified by small $\pm 1^{\circ} C $ temperature fluctuations occurring during day to night transitions at Cerro Paranal. The objective with this requirement is to prevent mirror daily calibration. The long-term mount stability describes mount stability for long periods of time, over a range of $\pm 10^{\circ} C $ with $1^{\circ} C $ variation per hour. This results in a total cycle duration of 20 hours, representing worst-case daily fluctuations.

FM2 has the particularity to have a different configuration for all 16 of them, as seen in Figure \ref{fig:FM2_configs}. There are a total of 8 angles, defined to fold each light path within the instrument. These 8 configurations are then found symmetric on the other side of the instrument. For each of these configurations, the requirements stay identical.

\subsection{FM2 Mechanical Designs}
\label{sec:mechanics}

\begin{figure}[!h]
    \centering
    \begin{subfigure}[b]{0.45\textwidth}
        \includegraphics[width=\textwidth]{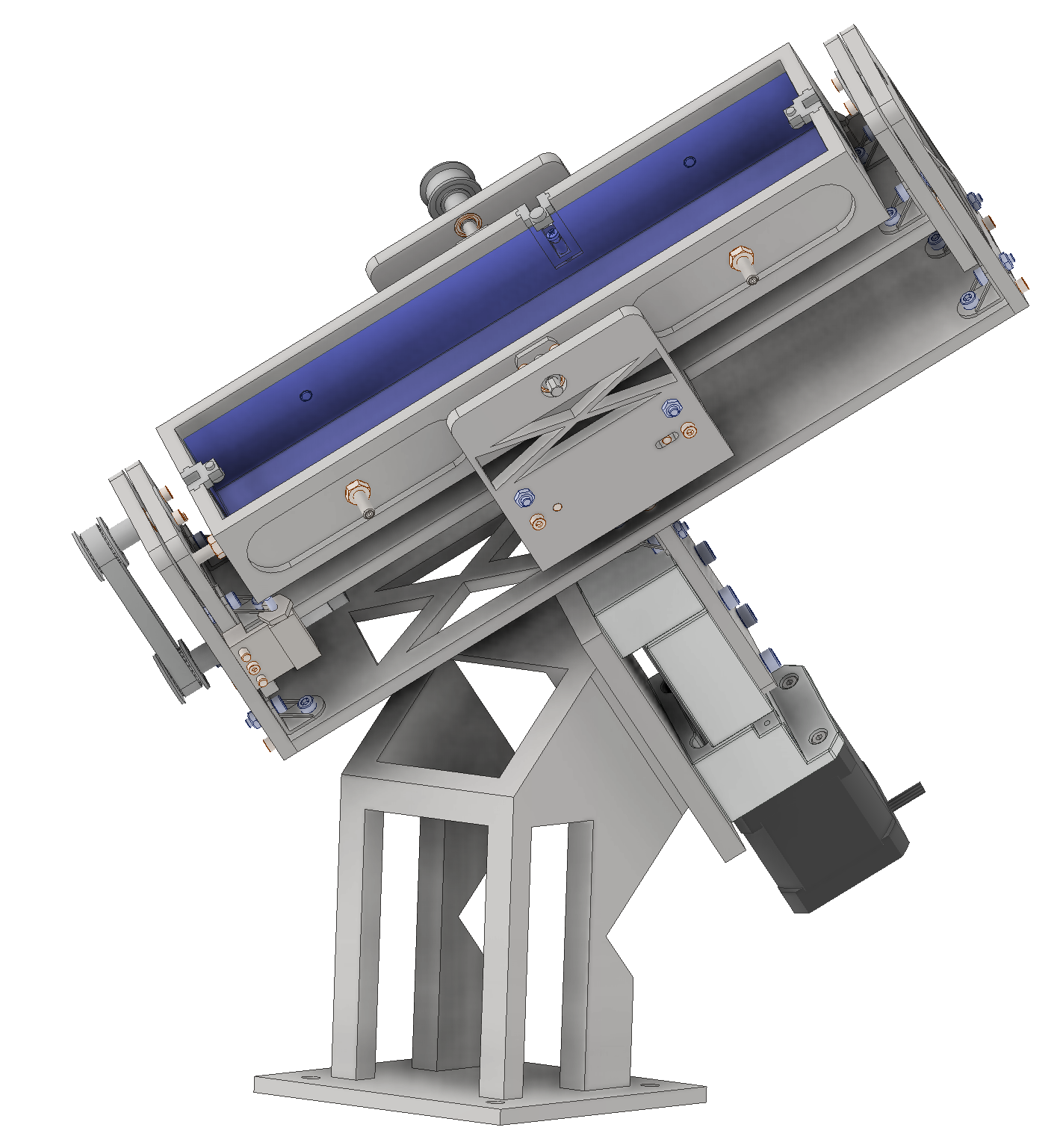}
        \caption{}  
        \label{fig:FM2_Evan}
    \end{subfigure}
    \hfill
    \begin{subfigure}[b]{0.45\textwidth}
        \centering
        \includegraphics[width=\textwidth]{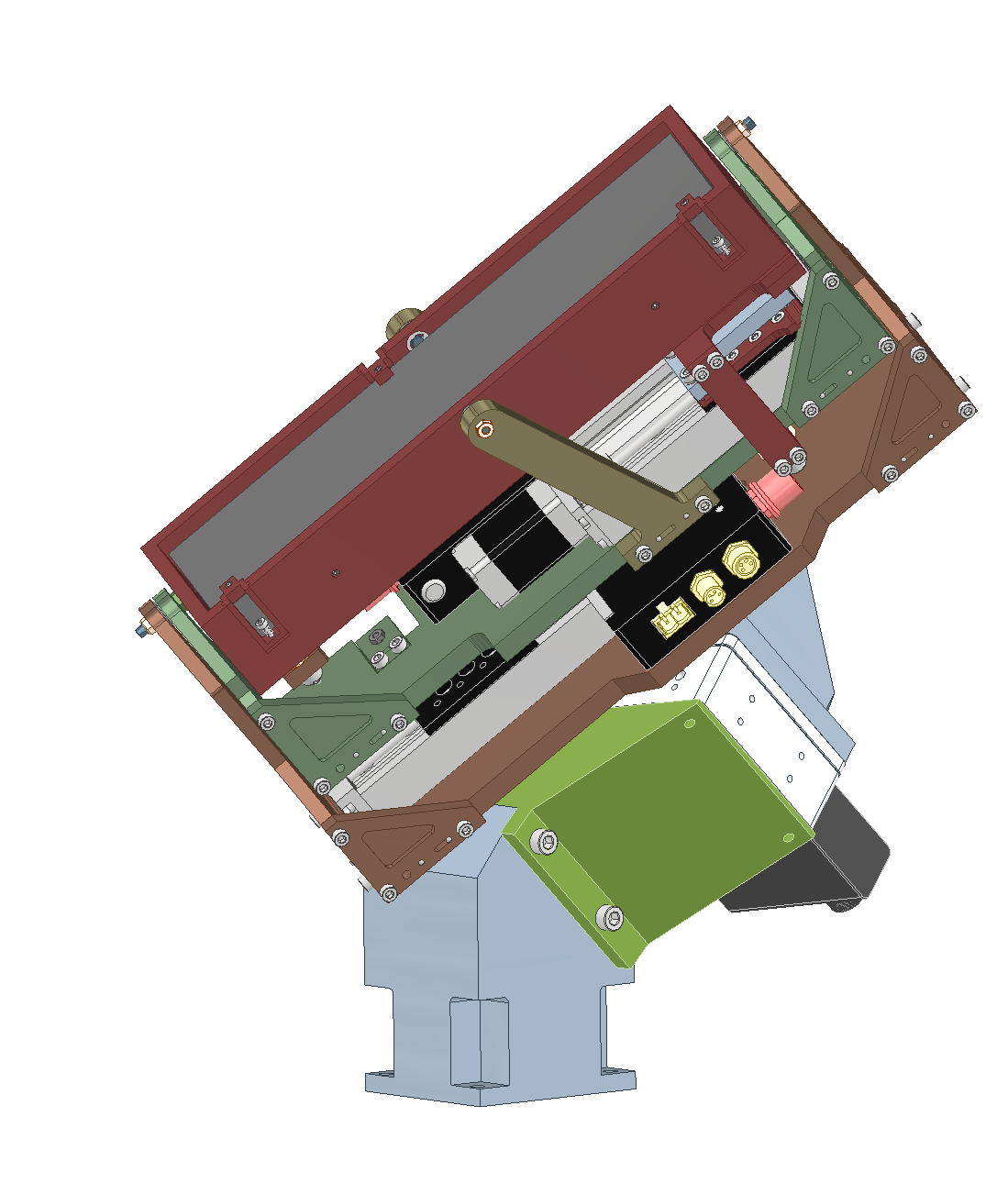}
        \caption{}  
        \label{fig:FM2_Gloria}
    \end{subfigure}
    \caption{Folding mirror 2 (FM2) designs: (a) Pulley-based gimbal design of FM2; (b) Slope and ball contact based design of FM2.} 
    \label{fig:FM2_new}
\end{figure}

Two FM2 designs, shown in Figure \ref{fig:FM2_new}, were developed and compared to evaluate their ability to meet the requirements listed in Table \ref{tab:FM2_req}.

The first design investigated for FM2 is a gimbal-based architecture incorporating two nested rotational stages to provide the required $\Theta X_{FM}$ and $\Theta Y_{FM}$ degrees of freedom. The mirror is mounted above a gimbal assembly composed of two perpendicular pivot axes, resulting in a decoupled kinematic structure. Each rotational axis is actuated by a piezoelectric motor driving a pulley-belt transmission, allowing remote motor placement within the constrained instrument envelope and maintaining a compact design. Vertical translation is provided by a compact linear stage mounted beneath the gimbal support plate, while the configuration's mirror orientation is defined by the geometry of the base structure.

Brushless DC motors were initially investigated for this design. However, due to the high mechanical transmission ratio between motor rotation and mirror displacement, they could not reach the desired repeatability requirement. Thus, piezo motors were considered, as they can reach 0.2 arcseconds of resolution. However, the encoders for these motors have a precision of 400 arcseconds, making it difficult to attain the desired performance for the mount. To be compliant, this design would need additional mechanical reduction, if the motor step cannot be reached in practice.

The main advantages of this architecture are its simple kinematics and straightforward control strategy. The decoupling of the two rotational axes simplifies calibration and minimizes cross-coupling between motions.

However, the use of a belt transmission raises concerns regarding motion reduction and long-term thermo-mechanical stability. The theoretical thermal expansion of the neoprene belt can be estimated using

\begin{equation}
\Delta L = \alpha L_0 \Delta T,
\end{equation}

where $\alpha$ is the coefficient of thermal expansion of neoprene, $L_0$ is the nominal belt length, and $\Delta T$ is the temperature variation. Using the values selected for the preliminary design ($\alpha = 2\times10^{-4}\,\mathrm{K^{-1}}$, $L_0 = 160~\mathrm{mm}$, and $\Delta T = 20^\circ\mathrm{C}$), the resulting belt length variation is

\begin{equation}
\Delta L = 0.64~\mathrm{mm}.
\end{equation}

Although a spring-loaded idler maintains belt tension and prevents slack, differential thermal expansion between the belt and the metallic support structure modifies the belt preload and transmission stiffness. Consequently, the effective transmission characteristics may vary with temperature, potentially generating small angular offsets and repeatability errors that are difficult to compensate in the required open-loop operation. In addition, the belt introduces potential long-term effects such as creep, aging, and wear. For these reasons, despite its mechanical simplicity and decoupled architecture, this design was considered to present an increased risk with respect to the thermal stability requirements of the FM2 mechanism.

\begin{figure}[!h]
    \centering
    \begin{subfigure}[b]{0.8\textwidth}
        \centering
        \includegraphics[width=\textwidth]{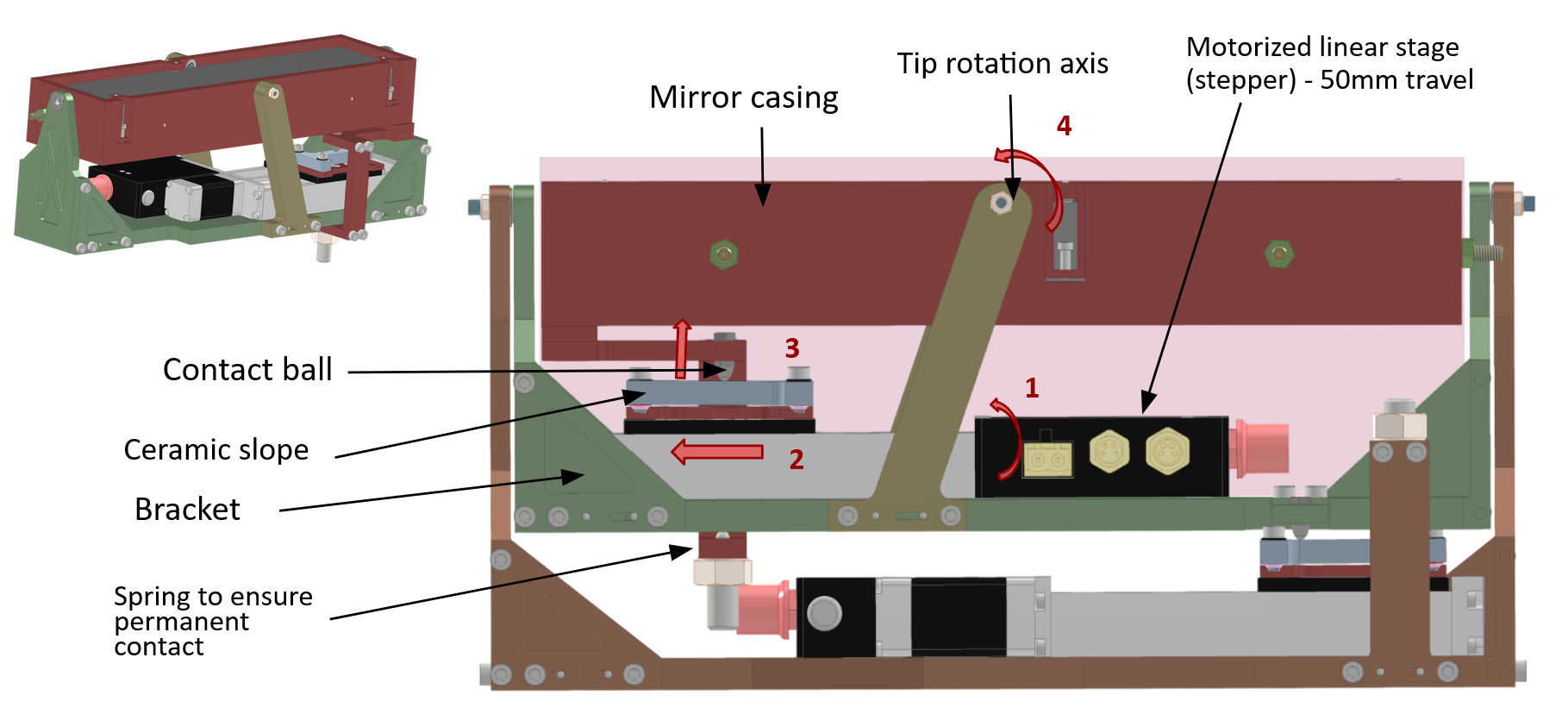}
        \caption{}
        \label{fig:FM2_tiptilt}
    \end{subfigure}
    \hfill
    \begin{subfigure}[b]{0.8\textwidth}
        \centering
        \includegraphics[width=\textwidth]{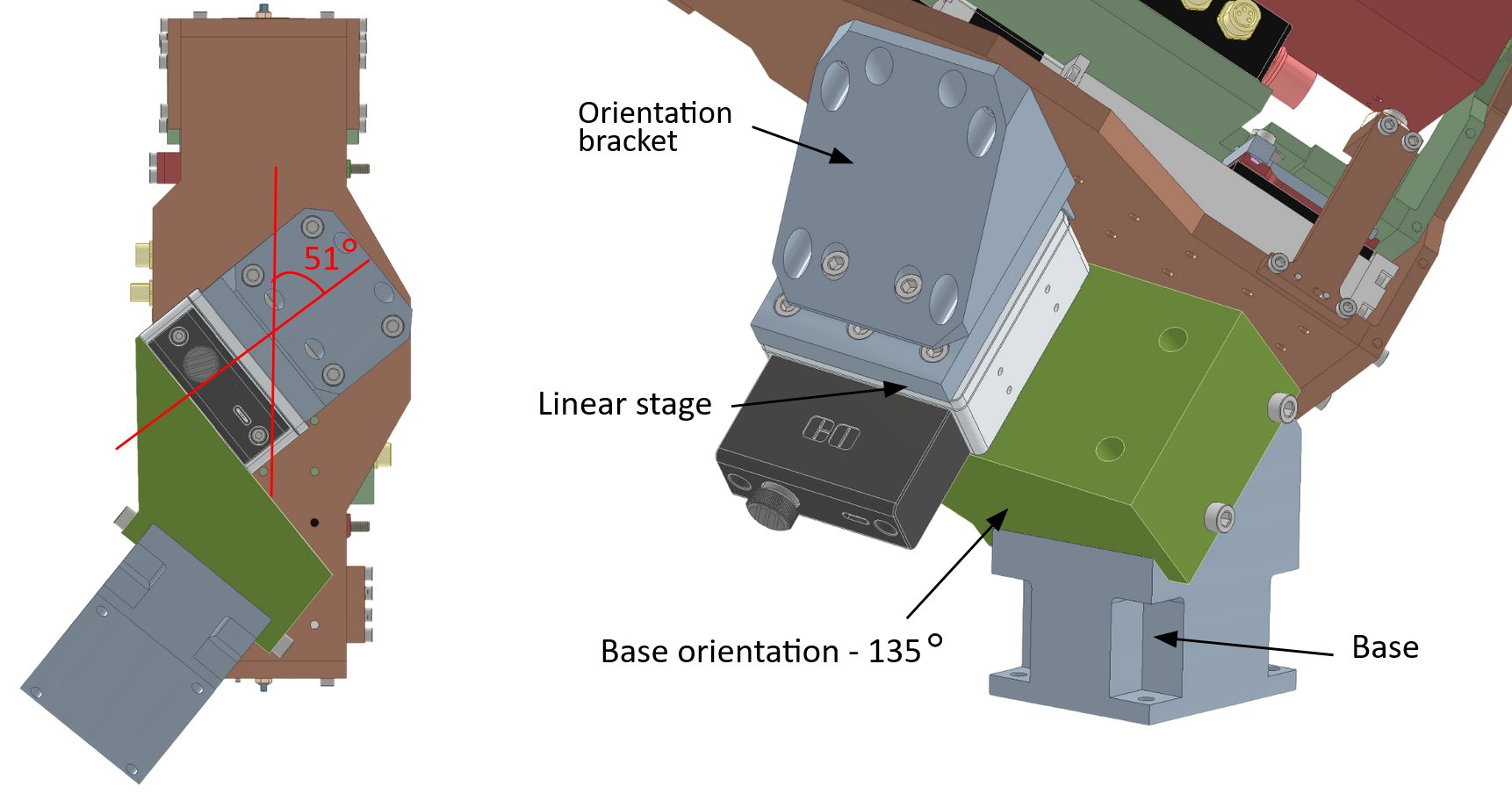}
        \caption{}
        \label{fig:FM2_z_mech}
    \end{subfigure}
    \caption{FM2 high motion reduction design: (a) Side view of FM2 showing the motion transmission for the $\Theta Y_{FM}$ rotation; (b) Bottom view of FM2 showing the motion transmission for the $Z_{FM}$ translation.}
    \label{fig:FM2_z}
\end{figure}

The second design, shown in Figure \ref{fig:FM2_Gloria}, is based on a ball-and-slope mechanism that provides a large motion reduction between the motor rotation and the resulting angular displacement of the FM2 mirror. This reduction minimizes the influence of actuator resolution, gearbox backlash, manufacturing tolerances, and mechanical play on the final mirror position.

This mechanism is derived from the FM1 architecture \cite{Mellinand2026FM1}, where a ball contact is used to achieve high-resolution angular positioning for the $\Theta X_{FM}$ and $\Theta Y_{FM}$ motions. Due to the limited space available around the mirror, the complete mechanism is positioned beneath the mirror and integrated within a gimbal structure. Both rotational axes are aligned with the mirror surface, as illustrated by the location of the pivot points.

The system is actuated by a DC motor integrated within a linear stage. The motor rotates a lead screw, guided by two rods (1 in Figure \ref{fig:FM2_tiptilt}), which drives a linear carriage (2) over a stroke of 50 mm. Mounted on this carriage is a high-stiffness EKasic ceramic slope designed to withstand bending under the applied contact loads. The slope is inclined using a thin shim over a length of 64 mm. A ball bearing is pressed against the slope by a preload spring, ensuring continuous contact throughout the earthquake acceleration range specified for the instrument. As the carriage moves, the ball undergoes a vertical displacement (3), which is transmitted to a lever arm directly attached to the mirror housing, inducing a rotation about the $\Theta X_{FM}$ axis (4).

The mirror itself is mounted within a dedicated housing and constrained using spring-loaded contacts to provide an isostatic mounting configuration. This arrangement minimizes the risk of parasitic motions affecting the alignment of the reflected beam.

The mechanism provides a theoretical transmission ratio of

\[
\eta = \frac{\theta_{mirror}}{\theta_{motor}} = 0.00172~\mathrm{arcsec/rad},
\]

where $\theta_{motor}$ is the motor rotation before the gearbox and $\theta_{mirror}$ is the resulting mirror rotation. This high reduction ratio results from the combination of the ball-and-slope mechanism, the lead-screw-driven linear stage, and the motor's 231:1 reduction gearbox. The resulting motion reduction enables fine angular positioning. The current geometry provides a theoretical angular range of approximately $6~\mathrm{arcmin}$, compatible with the required $5~\mathrm{arcmin}$ range. Furthermore, the slope angle can be adjusted by modifying the shim thickness, allowing the transmission ratio and angular range to be tuned if required.

The same operating principle is applied to the $\Theta Y_{FM}$ rotation stage. In this case, the pivots are positioned on the outer sides of the mirror assembly and reinforced using brackets to improve structural stiffness.

The vertical translation mechanism, shown in Figure \ref{fig:FM2_z}, is actuated at the base of the system using a compact linear stage. Unlike the rotational axes, no additional transmission mechanism is required. The selected stage provides a travel range of 15~mm, with a specified accuracy of 5~$\mu$m and repeatability of 2~$\mu$m, thereby exceeding the FM2 requirements. The stage is mounted directly at the configuration angle, allowing different optical-path orientations to be accommodated by modifying the attachment interface while maintaining a common base structure.

This design benefits from the high motion reduction provided by the ball-contact mechanism, making it inherently less sensitive to actuator resolution, gearbox backlash, and thermal variations. However, the architecture remains mechanically complex and requires numerous structural components to ensure sufficient stiffness. Future design iterations will focus on reducing the dimensions of the mechanism, reducing unused volume, and shortening lever arms to further increase stiffness and compactness.

\subsection{Electronics and Software Architecture}
\label{sec:electronics_software}

\begin{figure}[!h]
    \centering
    \includegraphics[width=\textwidth]{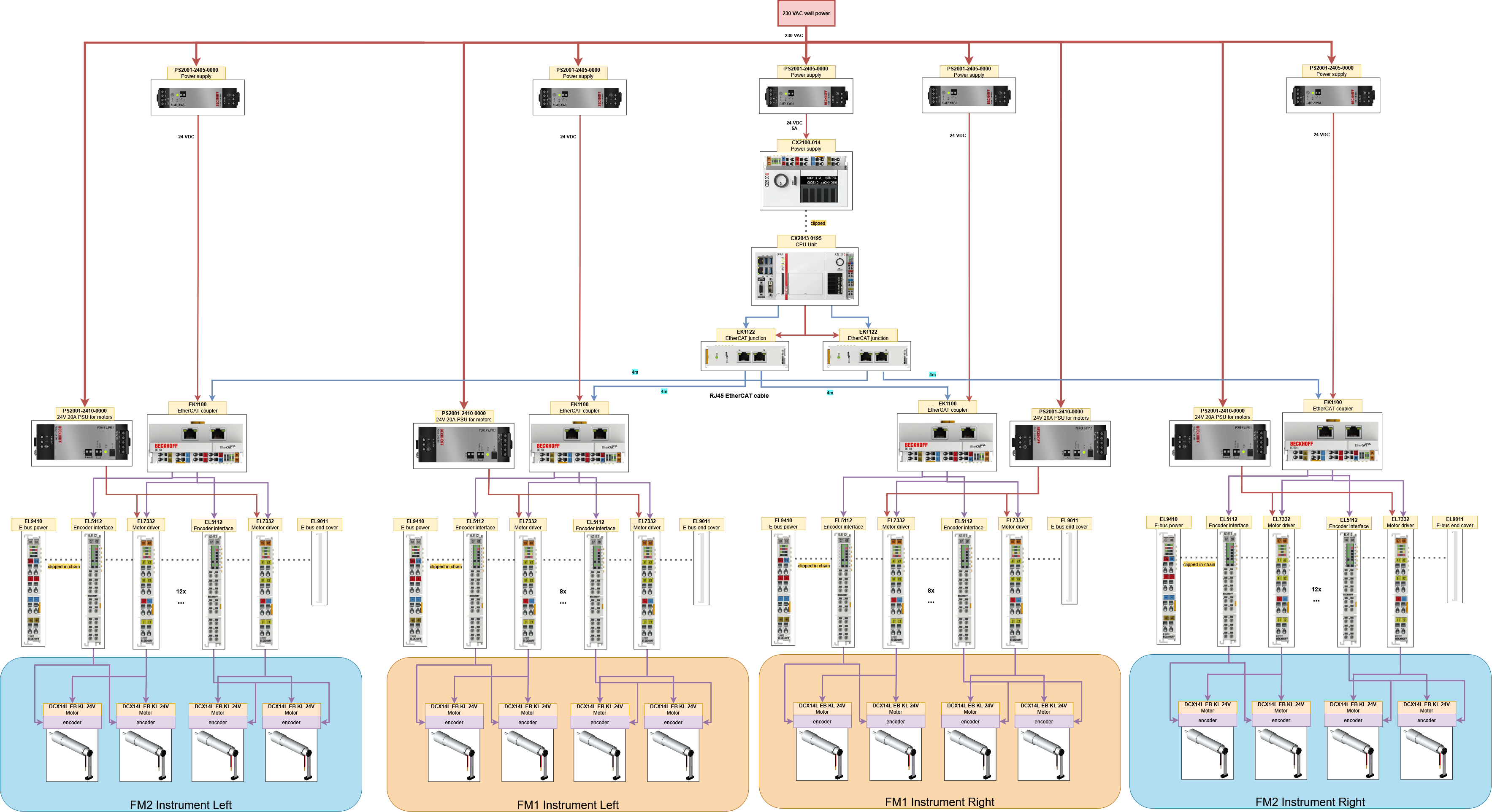}
    \caption{BlueMUSE folding mirror electronics architecture based on the Beckhoff EtherCAT environment.}
    \label{fig:Beckhoff}
\end{figure}

The FM2 control architecture follows the same approach as that developed for the FM1 mechanism \cite{Mellinand2026FM1}. The system consists of three motorized axes corresponding to the $\Theta X_{FM}$, $\Theta Y_{FM}$ and $Z_{FM}$ degrees of freedom. Each axis is equipped with a DC motor, an encoder, and limit switches used for homing and reference position acquisition.

Within BlueMUSE, the folding mirror mechanisms are integrated into the Beckhoff EtherCAT environment shown in Figure \ref{fig:Beckhoff}. DC motor actuation is performed using EL7332 motor controller terminals, while encoder feedback is acquired through EL5112 interface terminals. The Beckhoff PLC executes the low-level axis control, homing procedures, and communication with the instrument-level control system. The EtherCAT network is distributed through four branches within the instrument, simplifying power distribution, data communication, and cable routing.

As for FM1, the FM2 mechanism is intended for occasional alignment corrections rather than continuous operation during observations. Therefore, an open-loop control strategy is adopted. Each movement is initiated from a homed reference position in order to minimize the accumulation of positioning errors arising from backlash or mechanical compliance. The target mirror position is then reached using a calibration look-up table that accounts for non-linearities in the mechanical transmission.

The software architecture is organized into three hierarchical layers. The first layer provides low-level motor control functions, including current, velocity, and position regulation. The second layer manages axis-level functions such as homing, limit-switch monitoring, and conversion between mirror coordinates and motor positions. The third layer interfaces with the BlueMUSE control system, receiving alignment commands and issuing the corresponding position requests to the FM2 axes.

\section{Experimental Validation of FM2 Thermo-Mechanical Stability}

A dedicated optical metrology setup will be developed to evaluate the thermo-mechanical stability of the FM2 assembly under operating conditions. The objective of the experiment is to quantify thermally induced angular variations of the mirror with sub-arcsecond precision and verify compliance with the stability requirements summarized in Table \ref{tab:FM2_req}. 

\begin{figure}[!h]
    \centering
    \includegraphics[width=\textwidth]{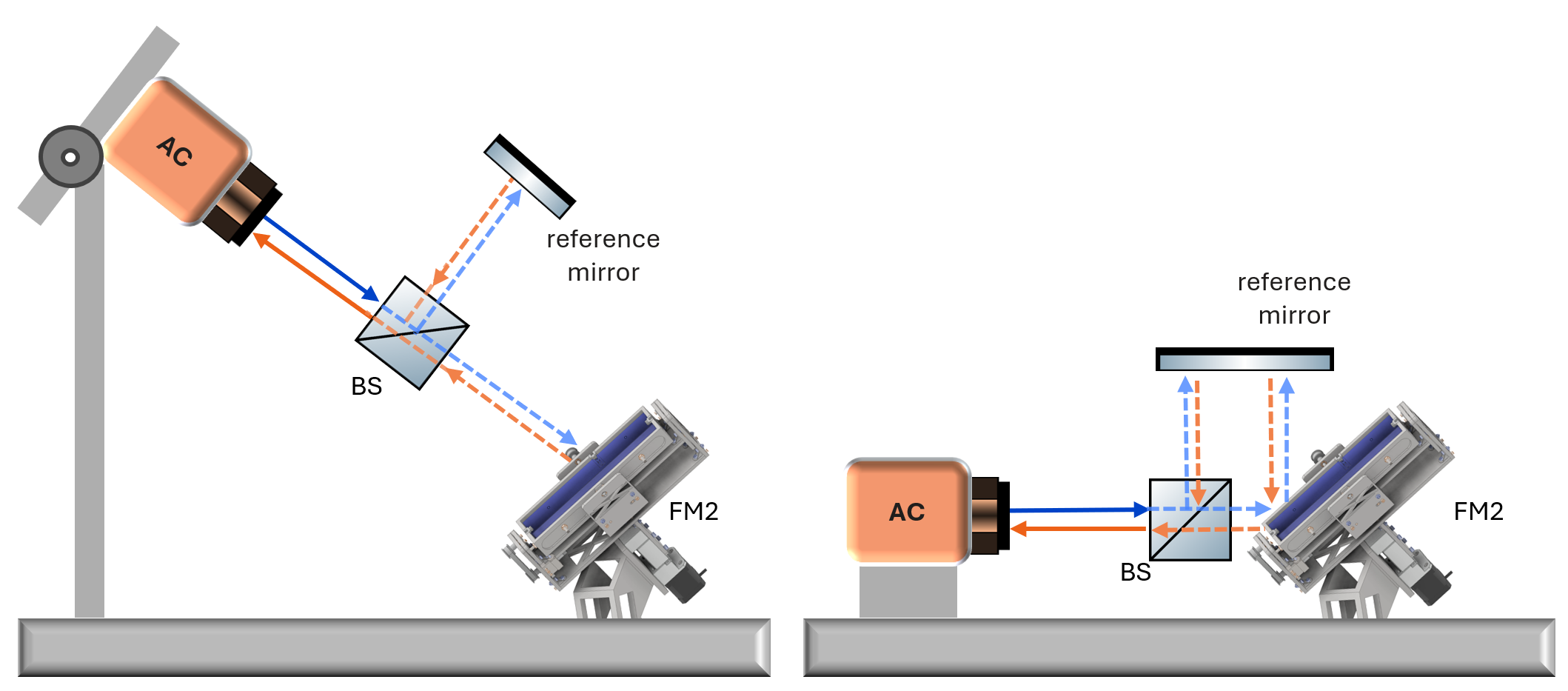}
     \begin{subfigure}[a]{0.40\textwidth}
        \caption{}  
        \label{fig:FM2_configs_a}
    \end{subfigure}
    \hfill
    \begin{subfigure}[b]{0.55\textwidth}
        \centering
        \caption{}  
        \label{fig:FM2_configs}
    \end{subfigure}
    \caption{Schematic representation of the optical metrology setup that will be used for FM2 thermo-mechanical stability measurements. A high-resolution autocollimator (AC) and a pellicle beam splitter (BS) provide simultaneous monitoring of the FM2 and a reference mirror, enabling differential angular measurements with sub-arcsecond precision. The AC will be either (a) mounted in respect to the main optical axis of the FM2 mirror and could be rotated accordingly or (b) mounted parallel to the optical table and use the reference mirror to redirect the AC beam back to FM2 and to the BS. This proposed experimental setups are still in the design phase.}
    \label{fig:Experimental setup}
\end{figure}

The proposed experimental setup is based on a high angular resolution electronic autocollimator (Figure \ref{fig:Experimental setup}). It will be either mounted above the optical bench - as shown in \ref{fig:FM2_configs_a} - so that the instrument will project a collimated beam directly into the FM2 mirror positioned at an incidence angle of one of its operational configurations shown in Figure \ref{fig:FM2_MUSE}. Variations in mirror orientation result in a corresponding displacement of the reflected beam on the autocollimator's detector, providing a direct measurement of angular motions. The other proposed solution is presented in \ref{fig:FM2_configs}, where the autocollimator is mounted horizontally in the optical table and its reflection to the reference - or another - mirror is sent back to the beam splitter. The proposed experimental setups need to be further evaluated or tested since they are still in the design phase. 

To distinguish mirror-induced motion from environmental and instrumental effects, the optical configuration shown schematically in Figure \ref{fig:Experimental setup} will be implemented. A pellicle beam splitter located at the autocollimator output divides the beam into two independent optical paths: one directed toward the FM2 and the other toward a reference mirror mounted on a mechanically stable support. The two reflected beams are simultaneously detected by the autocollimator, enabling simultaneous monitoring of the FM2 test and the reference signal. Any thermally induced angular displacement of the FM2 produces a lateral shift in the return signal. This shift is measured with sub-arcsecond precision. 

This dual-path architecture provides systematic disturbances that need to be quantified and compensated, improving the accuracy of the FM2 stability assessment. In particular, thermal expansion of the optical bench, mount relaxation, autocollimator drift and other environmental fluctuations. Since these effects influence both optical paths in a similar manner, differential measurements allow the true angular response of the FM2 assembly to be isolated with accuracy.

A pellicle beam splitter is selected over a conventional cube beam splitter due to its negligible optical path displacement, low level of ghost reflections and reduced sensitivity to temperature-induced refractive index variations. These characteristics minimize systematic measurement errors and enhance long-term measurement stability during extended thermal tests.

Repeatability tests will be performed without actuation in an ambient laboratory environment. The optical metrology setup will measure the differential angle between the reference path and FM2 while each axis repeatedly performs homing and targeting procedures. The resulting beam deviations at targets will then be analyzed to determine the repeatability of the mirror mount. 

Thermal stability measurements will be performed according to the environmental conditions defined by the BlueMUSE requirements. Short-term stability tests will evaluate the mirror response under temperature fluctuations of ±1 °C representative of day-to-night transitions at Cerro Paranal. Long-term stability tests will consist of thermal cycles of ±10 °C with a controlled variation rate of approximately 1 °C/h, corresponding to a total thermal cycle duration of approximately 20 h. Temporal drift of the measurement will be continuously tracked using the reference beam path, enabling compensation of the instrumental drift. 

During each thermal cycle, created by periodically cycling conventional heat flow using a fan and a heater, the autocollimator will continuously record the angular position of the reflected beams. The differential angular deviation between the FM2 and reference optical paths will be used to determine FM2's thermo-mechanical stability. The measured angular drift will then be compared with the specified stability requirement of 1.5 arcsec in both rotational degrees of freedom, providing a direct validation of the mount's suitability for long-term operation within the BlueMUSE instrument.

\section{Thermo-mechanical Analysis of the FM2 Mechanism}

Finite Element Analysis (FEA) was performed in ANSYS Workbench to evaluate the thermo-mechanical stability of the FM2 design under representative BlueMUSE operating conditions. The objective of the study was to quantify the influence of temperature variations on mirror alignment and to compare two candidate structural materials: stainless steel and Invar 36.

\subsection{Simulation Methodology}
\label{methodology}
The thermo-mechanical analysis was performed in two sequential steps. First, a steady-state thermal simulation was conducted by applying the representative thermal boundary conditions to the FM2 assembly. The resulting temperature field was then set as an input to the static structural analysis to compute thermally induced deformations and stresses.

The simulations focused on the complete FM2 mechanism, including the mirror support structure, the mirror itself and the high stiffness slopes bearing the mirror load. The results of the thermo-mechanical deformation of the mirror surface were then processed to determine their effect on the resulting variation of optical beam alignment.

Two structural materials were investigated for the mirror mount structure:

\begin{table}[!h]
\centering
\begin{tabular}{lcc}
\hline\hline
Property & FM2 in Stainless Steel & FM2 in Invar 36 \\
\hline
Density [g/cm$^3$] & 8.0 & 8.05 \\
Young's Modulus [GPa] & 193 & 141 \\
Poisson Ratio [-] & 0.30 & 0.29 \\
CTE [$10^{-6}$ K$^{-1}$] & 17.3 & 1.2 \\
Thermal Conductivity [W m$^{-1}$ K$^{-1}$] & 16 & 13 \\
Specific Heat [J g$^{-1}$ K$^{-1}$] & 0.50 & 0.51 \\
\hline
\end{tabular}
\caption{Material properties of the FM2 mount used for the thermo-mechanical simulations.}
\label{tab:FM2_materials}
\end{table}

The primary motivation for investigating Invar 36 is its very low coefficient of thermal expansion, which is expected to reduce thermally induced mirror motions resulting from the accumulation of thermal deformations throughout the mechanical structure. However, Invar 36 is significantly more expensive than stainless steel and presents additional manufacturing challenges due to its rapid work-hardening. The mirror material assigned is Zerodur, used for its extremely low coefficient of thermal expansion, and the slopes are defined as EKasic silicon carbide.

The thermal loading was defined to be representative of the day-to-night temperature variations expected at Cerro Paranal. The objective was to reproduce the gradual temperature decrease experienced by both the instrument structure and the surrounding air during evening operation. The simulations were performed for a total temperature variation of $4^{\circ}$C over a period of 10 hours, representative of the short term thermal stability requirement.

A weak natural convection boundary condition of $5~W\times m^{-2}\times °C^{-1}$ was applied to all exposed surfaces of the assembly. The ambient convection temperature was defined as a time-dependent function:

$T_{conv}[^\circ C] = 20 - 0.6 \times 1.1 \times 10^{-4} \times time[s]$

corresponding to a gradual decrease from $20^\circ$C to $17^\circ$C over 10 hours.

The mounting interface at the base of the FM2 was assumed to cool more rapidly than the surrounding air due to its thermal connection with the instrument structure. Thus, a second time-dependent temperature boundary condition was applied at the base:

$T_{base}[^\circ C] = 20 - 1.1 \times 10^{-4}\times time[s]$

corresponding to a decrease from $20^\circ$C to $16^\circ$C over 10 hours.

The resulting transient temperature field was then transferred to the structural analysis, where both thermal loads and gravitational loading were applied in order to evaluate the resulting deformations and stress distributions.

\subsection{Transient Thermal FM2 Analysis}

\begin{figure}[!h]
    \centering
    \begin{subfigure}[b]{0.6\textwidth}
        \centering
        \includegraphics[width=\textwidth]{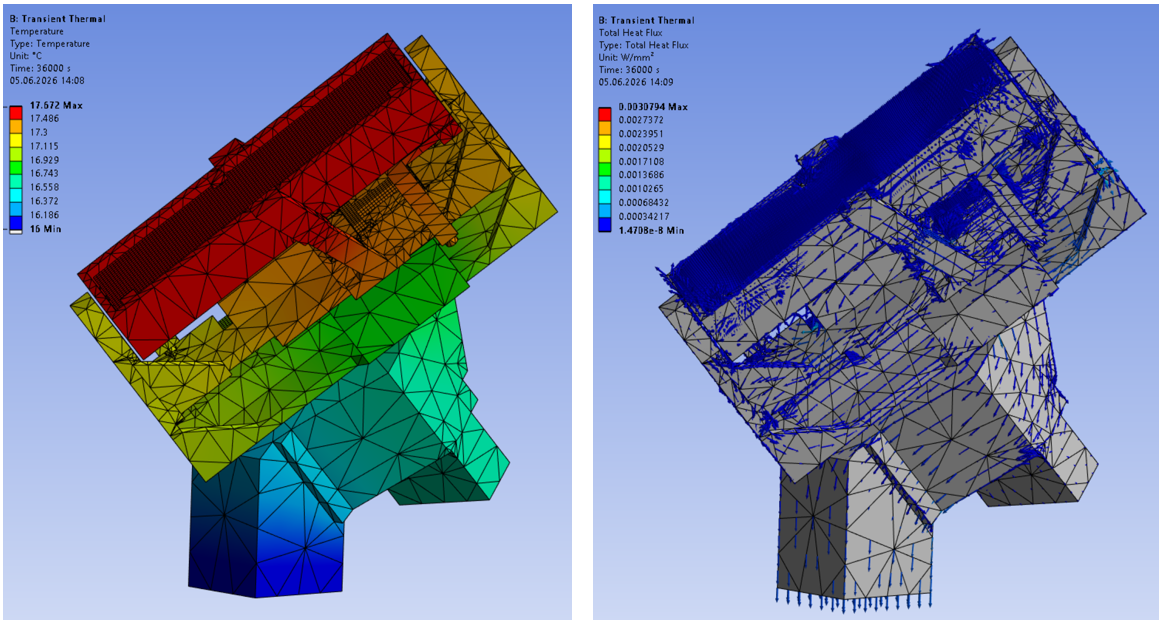}
        \caption{}
        \label{fig:thermal_ansys_steel}
    \end{subfigure}
    \hfill
    \begin{subfigure}[b]{0.6\textwidth}
        \centering
        \includegraphics[width=\textwidth]{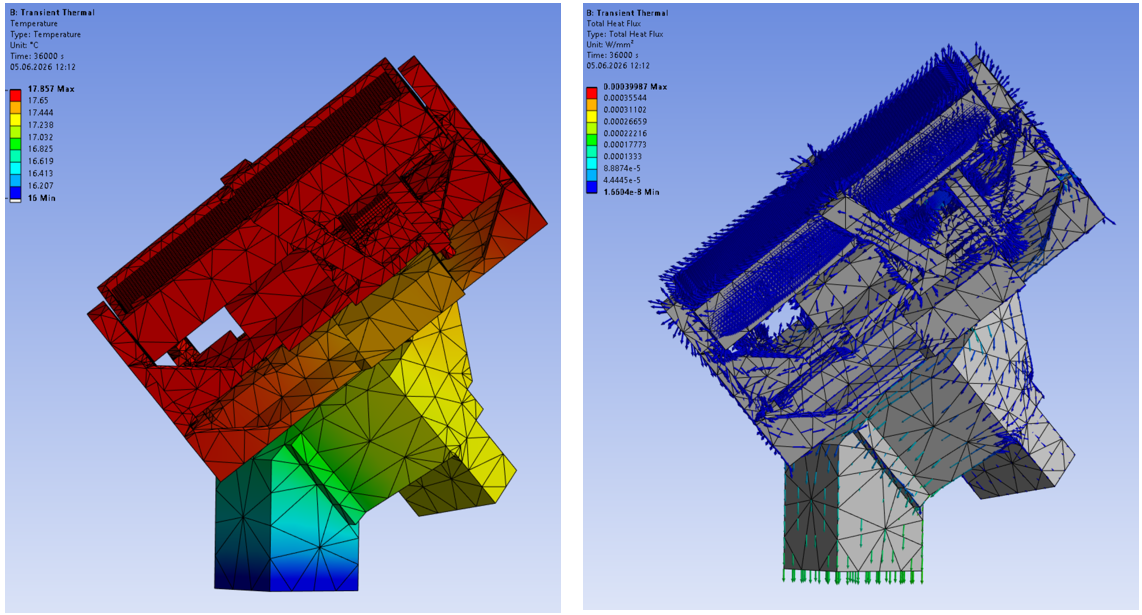}
        \caption{}
        \label{fig:thermal_ansys_invar}
    \end{subfigure}
    \caption{Temperature and heat flux analysis of the FM2 mirror mount. (a) Stainless steel configuration (b) Invar configuration.}
    \label{fig:FM2_z}
\end{figure}

Figures \ref{fig:thermal_ansys_steel} and \ref{fig:thermal_ansys_invar} show the transient temperature distribution and heat-flux evolution obtained for the FM2 assembly manufactured from stainless steel and Invar 36, respectively.

For both materials, the applied thermal boundary conditions generate a temperature gradient between the mounting interface and the mirror assembly. After 10 hours, the base of the mechanism reaches $16^\circ$C due to its direct thermal connection to the instrument structure, while the mirror remains at $17.8^\circ$C, following the warmer ambient air enclosed within the instrument.

The largest temperature gradients are observed around the ball-and-slope interfaces and at transitions between different materials, where thermal resistance slows heat propagation through the mechanism. While the steel FM2 experiences thermal gradients on the long pivot arms, this effect is attenuated on Invar where gradients occur around the base of the mechanism. This suggests that thermal deformations associated with the pivot arms may be reduced in the Invar configuration. Nevertheless, the temperature field remains relatively smooth throughout the structure, with no significant thermal concentration zones.

The corresponding heat-flux distributions show the primary thermal paths through the assembly during cooling. For both materials, the highest heat fluxes occur near the lower part of the gimbal structure connecting the mirror assembly to the base.

In both cases, the non-uniform cooling of the mechanism generates differential thermal expansion throughout the kinematic chain. These differential deformations can induce mirror rotations and affect the alignment of the reflected optical beam, motivating the following structural analysis.

\subsection{Mechanical FM2 Analysis}

\begin{figure}[!h]
    \centering
    \begin{subfigure}[b]{0.6\textwidth}
        \centering
        \includegraphics[width=\textwidth]{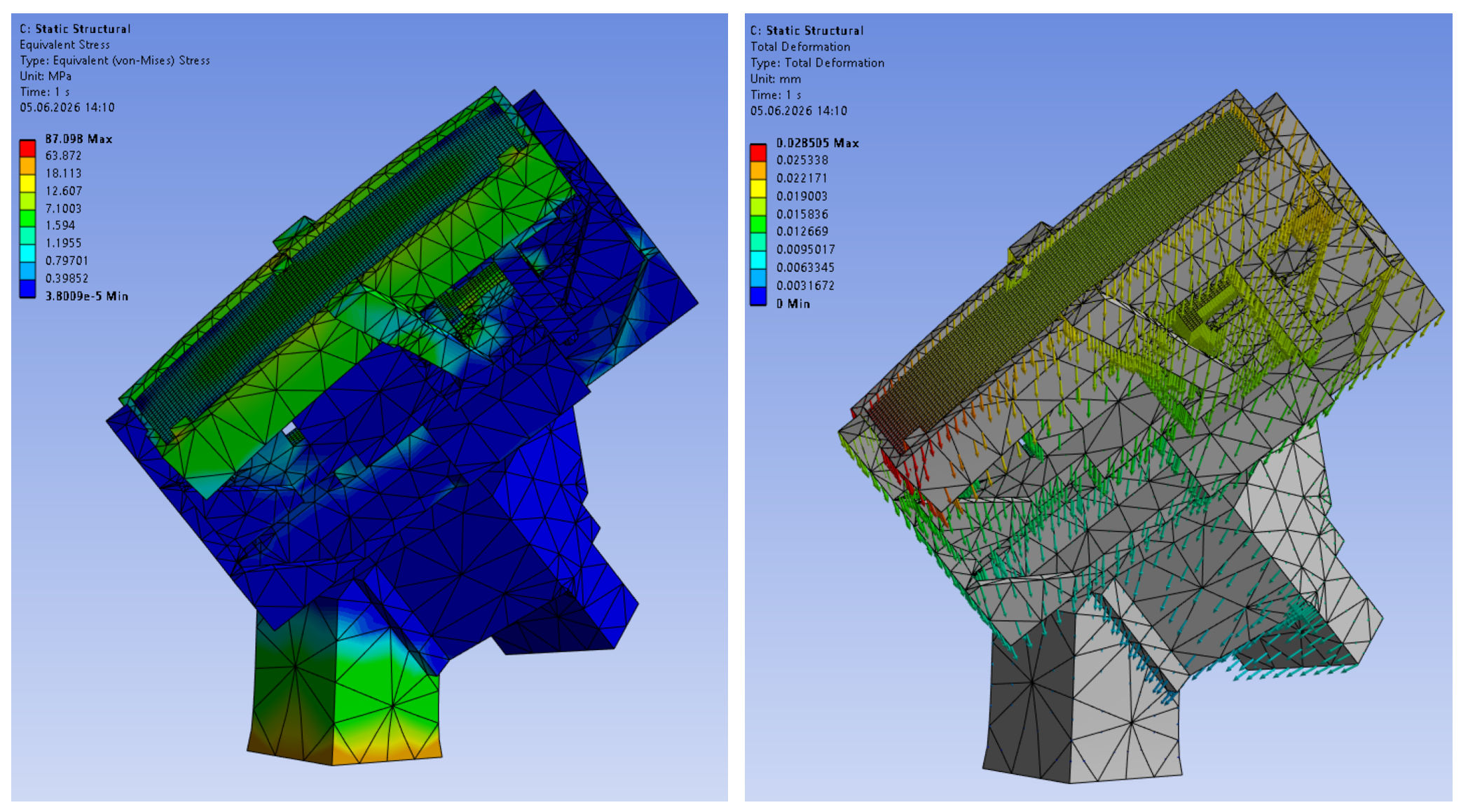}
        \caption{}
        \label{fig:mech_ansys_steel}
    \end{subfigure}
    \hfill
    \begin{subfigure}[b]{0.6\textwidth}
        \centering
        \includegraphics[width=\textwidth]{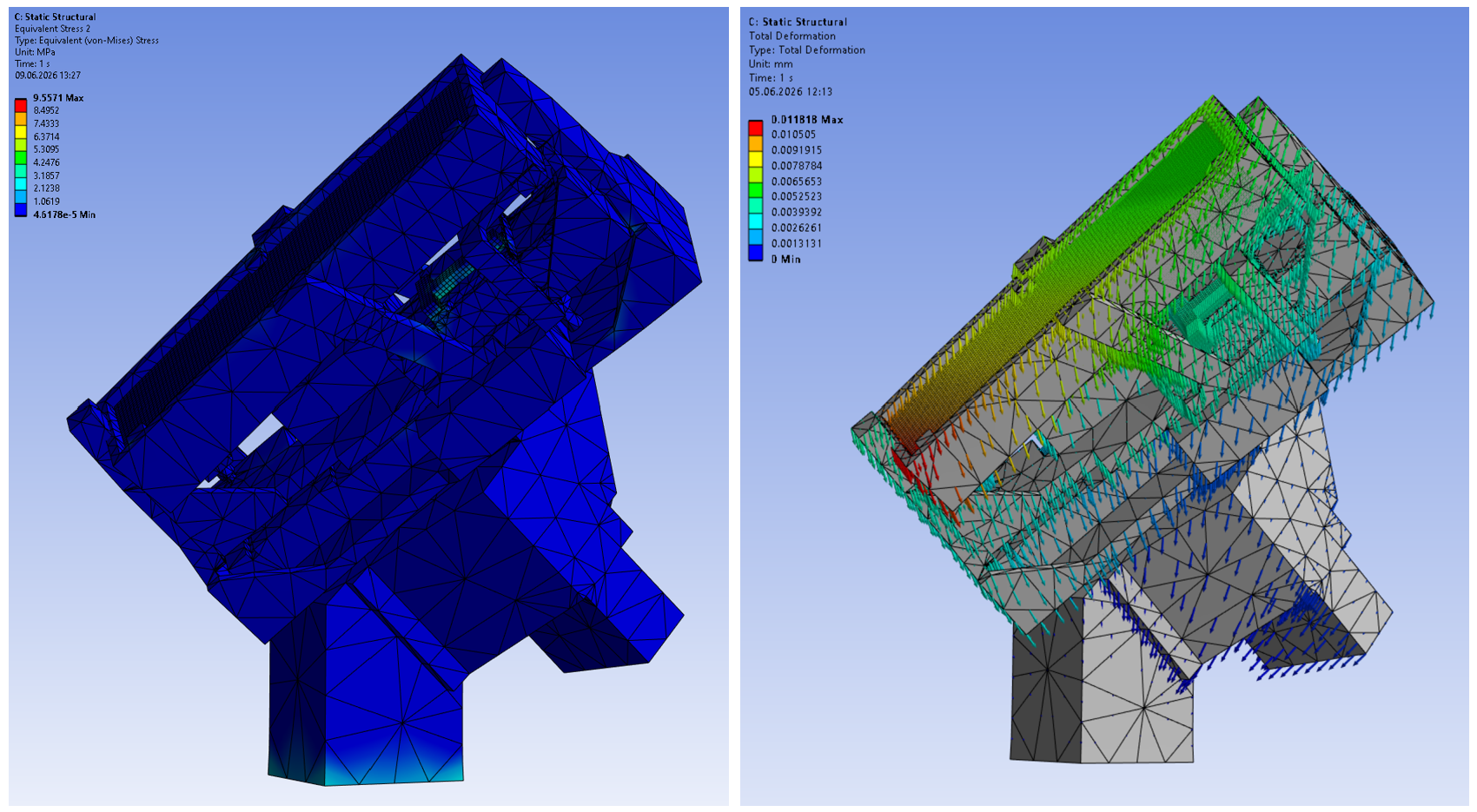}
        \caption{}
        \label{fig:mech_ansys_invar}
    \end{subfigure}
    \caption{Stress and total deformation of the FM2 mirror mount. (a) Stainless steel configuration (b) Invar configuration.}
    \label{fig:FM2_z}
\end{figure}

Figures \ref{fig:mech_ansys_steel} and \ref{fig:mech_ansys_invar} present the equivalent stress and total deformation resulting from the combined gravity and thermal loads described in section \ref{methodology}.

The maximum equivalent stress observed in the stainless-steel FM2 configuration is 87 MPa, while the Invar 36 design reaches 9.5 MPa only. The higher stress observed in the stainless-steel configuration is primarily caused by differential thermal expansion between bonded structural components. As the structure cools, the larger coefficient of thermal expansion of stainless steel creates greater displacements between the mirror support, the ceramic slope interfaces, and the optical components. These differential deformations are constrained by the assembly interfaces, resulting in localized stress concentrations that propagate through the kinematic chain toward the mirror mount. This effect is significantly lower with Invar, due to its low CTE. This makes the material a more viable solution to avoid assembly-level displacements and elastic deformation. 

To mitigate these effects, preload springs are incorporated into the design, as described in Section \ref{sec:mechanics}. These springs provide compliance between the EKasic slopes, Zerodur and the rest of the assembly, allowing a portion of the differential thermal expansion to be absorbed elastically rather than being transferred towards the mirror.

A similar stress distribution is observed for the Invar 36 configuration. However, its lower coefficient of thermal expansion results in smaller differential displacements throughout the structure, thus reducing the resulting stress.

The deformation results show that the maximum displacement predicted for the Invar 36 design is approximately 11.8 $\mu$m, compared to 28.5 $\mu$m for the stainless-steel configuration. The largest displacements occur at the mirror assembly and is created by the accumulation of thermal deformations throughout the kinematic chain. In addition, the ball-and-slope mechanism supports the mirror from one side, which creates a lever-arm effect that amplifies small structural deformations into mirror motions. 

Although the absolute displacements remain relatively small, micron-level motions at the mirror support can translate into significant angular deviations of the reflected optical beam. Thus, the deformation of the mirror surface is further analyzed in the following section through a beam-deviation calculation.

\subsection{Mirror Angular Deviation Analysis}

The optical beam deviation resulting from the thermal deformation was evaluated by calculating the change in orientation of the mirror surface.

Four points located on the optical surface of the bloc mirror were sampled from the finite-element model before and after deformation. The surface normal vector was computed from these points for both the undeformed and deformed configurations. The angular difference between the two normal vectors provides the mirror rotation caused by thermal loading.

For small angular variations, the mirror rotation vector can be approximated by

$\boldsymbol{\omega} \approx \mathbf{n}\times\mathbf{n'}$

where $\mathbf{n}$ and $\mathbf{n'}$ are the undeformed and deformed surface normals respectively.

The rotational components were then projected onto the local mirror axes corresponding to the $\Theta X_{FM}$ and $\Theta Y_{FM}$ degrees of freedom. Because reflection doubles angular errors, the reflected beam deviation is given by

$\theta_{beam}=2\theta_{mirror}$

A MATLAB post-processing script was developed to compute the mirror rotations and normal translation with corresponding beam deviations from the ANSYS deformation results.

The predicted beam deviation for the stainless-steel design reached 16.2 arcsec in $\Theta X_{FM}$ and 11.5 arcsec in $\Theta Y_{FM}$. For the Invar 36 configuration, the corresponding values were reduced to 14.5 arcsec and 10.5 arcsec respectively. As for the $Z_{FM}$ translation, the stainless-steel design provides a 12 $\mu m$ displacement, while the Invar has a 3.5 $\mu m$ displacement

These results indicate that the beam deviation is influenced not only by the thermal expansion of the structural material, but also by the mechanical architecture of the FM2 assembly and the geometric constraints imposed by the available space envelope. Differential thermal displacements accumulate throughout the kinematic chain and are transmitted directly to the mirror support, where they create mirror rotations. While the use of Invar significantly reduces the overall deformation of the structure, the reduction in beam deviation is smaller because the angular error depends primarily on the relative displacement between different regions of the mirror support rather than on the absolute displacement magnitude.

The deformation simulation suggest that the current mirror support arrangement is a significant contributor to the observed angular deviations. In particular, the mirror is primarily supported on one side through the ceramic slope interface, which can create a lever-arm effect that amplifies mirror rotations. In order to reach the required 1.5 arcsec stability, a future design iteration will investigate the addition of a guiding support on the opposite side of the mirror in order to improve load distribution and reduce thermally induced rotations and thermal links could be added to reduce gradients inside the mount.

\subsection{Discussion}

The thermo-mechanical analysis provides insight into the behaviour of the proposed FM2 architecture and identifies the primary contributors to mirror displacement under representative operating conditions.

The results confirm the benefit of using Invar 36 for the structural components of the mechanism. Compared to stainless steel, the Invar configuration reduces the overall deformation, stress, and mirror translation. In particular, the predicted normal translation decreases from approximately $12~\mu$m to $3.5~\mu$m, allowing compliance with the translational stability requirement, listed in Table \ref{tab:FM2_req}. These results demonstrate that selecting Invar as the mount material plays an important role in limiting thermally induced deformations within the assembly. Additionally, attaching thermal links on the mount to facilitate heat propagation throughout the structure can help minimize thermal deformations located on critical pivot points of the mirror.

The reduction in angular beam deviation is smaller. However, the simulations indicate that the dominant contribution to the observed mirror rotations originates from the current mirror support of the $\Theta Y_{FM}$ rather than from the thermal expansion of the structural material itself. The ball-and-slope mechanism supports the mirror primarily from one side, creating a lever-arm effect that amplifies under gravitational load. Additional guidance on the opposite side of the mirror could substantially reduce this displacement while preserving the advantages of the ball-and-slope architecture. 

Overall, the analysis highlights the path for future improvements on the FM2 design, and justify a preferential choice to a mount made in Invar 36.

\section{Conclusion}

This paper presented the development of a motorized three-degree-of-freedom Folding Mirror 2 (FM2) mount for the BlueMUSE instrument. Following a comparative study of several concepts, a ball-and-slope architecture derived from the FM1 mechanism was selected due to its high motion reduction, which will reduce the mirror's sensitivity to actuator backlash and manufacturing tolerancing.

The mechanical, electronics, and software architectures required for integration within the BlueMUSE Beckhoff EtherCAT environment were defined. The proposed control strategy relies on homing procedures and open-loop positioning using calibration tables, which is compatible with the expected operating mode of the folding mirrors, where only occasional alignment corrections are required.

A dedicated optical metrology setup based on a high-resolution autocollimator and a reference optical path was also developed to characterize the thermo-mechanical stability of the FM2 assembly. This test bench will enable differential angular measurements and provide the experimental validation necessary for future design iterations.

A first thermo-mechanical assessment of the design was performed using finite-element simulations representative of day-to-night temperature variations at Cerro Paranal. The study compared stainless steel and Invar 36 as candidate structural materials and evaluated their influence on temperature distribution, deformation, stress, and optical beam stability.

The results demonstrate the benefit of Invar 36 for reducing thermally induced deformation and stress within the mechanism. The predicted mirror translation satisfies the translational stability requirement, while the beam-deviation analysis identifies the mirror support system as the primary contributor to the angular displacement. 

Future work will therefore focus on refining the mirror support structure and the compactification the gimbal stages. Through this iterative approach, the FM2 mechanism is expected to achieve the stability and repeatability required for long-term operation within the BlueMUSE instrument.

\acknowledgments 
 
The authors would like to thank the Funding LArge international REsearch projects (FLARE) program for supporting the development of the FM1 motorized mount and the associated testbench instrumentation used for its characterization.

\bibliography{report} 
\bibliographystyle{spiebib} 

\end{document}